\documentclass[twocolumn]{aastex62}
\pdfoutput=1 
\usepackage{hyperref}
\usepackage{natbib}
\usepackage{appendix}
\graphicspath{{./}{figures/}}

\received{March 31, 2019}
\revised{June 17, 2019}
\accepted{July 2, 2019}
\submitjournal{ApJ}

\shorttitle{Stellar Local Arm}
\shortauthors{Miyachi et al.}

\begin{document}
\title{Stellar Overdensity in the Local Arm in $Gaia$ DR2}

\author{Yusuke Miyachi}
\affiliation{Department of Physics, Faculty of Science, Yamaguchi University, \\ Yoshida 1677-1, Yamaguchi-city, Yamaguchi 753-8512, Japan}

\author{Nobuyuki Sakai}
\affiliation{Mizusawa VLBI observatory, National Astronomical Observatory of Japan, \\
2-21-1 Osawa, Mitaka, Tokyo 181-8588, Japan}
\affiliation{Korea Astronomy $\&$ Space Science Institute, 776, Daedeokdae-ro, Yuseong-gu, Daejeon 34055, Korea}

\author[0000-0001-8993-101X]{Daisuke Kawata}
\affiliation{Mullard Space Science Laboratory, University College London, \\ Holmbury St. Mary, Dorking, Surrey RH5 6NT, UK}

\author[0000-0002-2154-8740]{Junichi Baba}
\affiliation{National Astronomical Observatory of Japan, \\
2-21-1 Osawa, Mitaka, Tokyo 181-8588, Japan}

\author{Mareki Honma}
\affiliation{Mizusawa VLBI observatory, National Astronomical Observatory of Japan, \\
2-21-1 Osawa, Mitaka, Tokyo 181-8588, Japan}
\affiliation{Mizusawa VLBI observatory, National Astronomical Observatory of Japan, \\
2-12 Hoshi-ga-oka-cho, Mizusawa-ku, Oshu, Iwate 023-0861, Japan}
\affiliation{The Graduate University for Advanced Studies (Sokendai),\\
Mitaka, Tokyo 181-8588, Japan}

\author{Noriyuki Matsunaga}
\affiliation{Department of Astronomy, The University of Tokyo,\\
7-3-1 Hongo, Bunkyo-ku, Tokyo 113-0033, Japan}



\author{Kenta Fujisawa}
\affiliation{Department of Physics, Faculty of Science, Yamaguchi University, \\ Yoshida 1677-1, Yamaguchi-city, Yamaguchi 753-8512, Japan}



\begin{abstract}

Using the cross-matched data of {\it Gaia}~DR2 and 2MASS Point Source Catalog,
we investigated the surface density distribution of stars aged {$\sim$}1~Gyr
in the thin disk in the range of $90^{\circ} \leq l \leq 270^{\circ}$.
We selected 4,654 stars above the turnoff corresponding to the age $\sim$1~Gyr, that fall within a small box region in the color--magnitude diagram, $(J-K_{\rm s})_0$ versus $M(K_{\rm s})$, for which the distance and reddening are corrected. The selected sample shows an arm-like overdensity at $90^{\circ} \leq l \leq 190^{\circ}$.
This overdensity is located close to the Local arm traced by
high-mass star forming regions (HMSFRs), but its pitch angle is slightly larger than that of the HMSFR-defined arm.
Although the significance of the overdensity we report is marginal,
its structure 
poses questions concerning both of the competing scenarios
of spiral arms,
the density-wave theory and the dynamic spiral arm model.
The offset between the arms traced by stars and HMSFRs, i.e., gas, is
difficult to be explained by the dynamic arm scenario.
On the other hand, the pitch angle of the stellar Local arm, if confirmed, 
larger than that of the Perseus arm is difficult to be explained
with the classical density-wave scenario. The dynamic arm scenario
can explain it if the Local arm is in a growing up phase,
while the Perseus arm is in a disrupting phase. Our result provides
a new and complex picture of the Galactic spiral arms,
and encourages further studies. 

\end{abstract}

\keywords{Galaxy: disk --- Galaxy: kinematics and dynamics --- Galaxy: solar neighborhood --- Galaxy: structure}


\section{Introduction} \label{sec:intro}



Revealing the shapes of spiral arms in the Milky Way is a long-standing challenge in Galactic astronomy \citep[e.g.][]{Vallee2017}. First successful identification of spiral arms in the Milky Way has been made by \citet{Morgan1952} from distributions of ionized hydrogen in the solar neighborhood. Since then, many studies have reported the characteristics of spiral arms in the Milky Way \citep[e.g.,][]{van_de_Hulst1954, Oort1958, Georgelin1976, Russeil2003, Paladini2004, Hou2014}. The well-known spiral arms within a few kiloparsecs of the Sun are the Sagittarius-Carina arm and the Perseus arm. The Sagittarius-Carina arm passes Galactic longitude of $l=0^{\circ}$ inside the solar radius, and the Perseus arm passes $l=180^{\circ}$ outside the solar radius. These features are revealed by a large number of gas and young stellar tracers, such as O and early B stars, giant molecular clouds and $\rm{H_{II}}$ regions \citep{Bok1970, Russeil2003, Hou2009, Hou2014, Monguio2015}. The Perseus arm is also associated with the excess of older stars \citep[][and references therein]{Churchwell2009}. On the other hand, such an excess of old stars has not been found around the Sagittarius arm \citep{Benjamin2005}. These lead to an ongoing debate that the Perseus arm could be one of two major spiral arms in the Milky Way, while the Sagittarius-Carina arm is a minor gaseous spiral arm \citep{Drimmel2000, Benjamin2008, Churchwell2009}. Spiral patterns in the Galactic plane are also traced by high-mass star-forming regions (HMSFRs) whose precise parallaxes can be measured using Very Long Baseline Interferometry \citep[VLBI;][]{Reid2009, Honma2012, Reid2014b}. More than 100 HMSFRs have been identified at the expected locations of the Sagittarius-Carina and Perseus arms \citep{Reid2014}.

There is another closest spiral arm between the Sagittarius-Carina and Perseus arms. This spiral arm is called the ``Local arm'' and also called the ``Orion Arm'' \citep{van_de_Hulst1954, Georgelin1976, Reid2014, Xu2016}. The Local arm is identified with the neutral hydrogen gas, $\rm{H_{II}}$ regions, Cepheids, OB stars and HMSFRs \citep{Walraven1958, Bok1959, Becker1970, Hou2014, Xu2016, Xu2018}. However, the features of the Local arm is not very clear, compared to the Sagittarius-Carina and Perseus arms. The Local arm is often considered to be a branch-like features or a spur which bridges between the Sagittarius-Carina arm and the Perseus arm \citep{Oort1952, Morgan1953, van_de_Hulst1954, Kerr1970, Kerr1970b, Russeil2003, Russeil2007, Xu2016, Xu2018}. This infers that the Local arm is not a major spiral arm, but a secondary minor spiral feature, which is just connected patchy star forming regions only traced by gas and very young stars \citep{Gum1955, Bok1959, Bok1970, Kerr1970, Georgelin1976, Hou2014}. However, a large number of HMSFRs are observed in the Local arm, and the overall length ($>$ 5~kpc) identified with HMSFRs is substantial, which led to a recent debate that the Local arm may be a major spiral arm \citep{Xu2013, Xu2016}. If the Local arm is a major spiral arm, we should be able to identify the stellar overdensity of the Local arm, which has not been observed, despite its being the closest spiral arm. 

The European Space Agency's {\it Gaia} mission \citep{GaiaDR1a} has made their second data release \citep[{\it Gaia}~DR2;][]{Gaia2018a}. {\it Gaia}~DR2 provides the position, parallax and proper motions for $\sim 1.3 \times 10^9$ stars in the Milky Way \citep{Lindegren2018}, and radial velocity for about 7 million stars \citep{Soubiran2018} measured with Radial Velocity Spectrograph \citep[RVS;][]{Cropper2018}. The precise measurement of the parallax for the bright stars around the Local arm in $Gaia$ DR2 enables us to study the stellar distribution for a selected population of stars. In this paper, to answer the question if or not there is a stellar arm associated with the Local arm, we map the stellar density of a specific population of stars with about 1~Gyr of age at $90<l<270^{\circ}$. The 1~Gyr population is chosen, because they are significantly older than the previously known Local arm tracers and they are bright and more uniquely located in the Hertzsprung$-$Russell (HR) diagram. Cross-matched sample of {\it Gaia}~DR2 stars with the Two Micron All Sky Survey Point Source Catalog \citep[2MASS PSC;][]{Skrutskie2006} are used to identify the population. We also evaluated the completeness of our selected sample against the 2MASS PSC, and confirmed that our sample has reasonable completeness within the distance of 0.2 and 1.3~kpc. 

If the stellar arm is identified, the positional offset between the gas and stellar arms, it would be interesting to consider the origin of the spiral arm \citep[][]{Dobbs2010,Baba2015,Egusa2017}. Recently, the origin of the spiral arms is hotly debated. There are two major competing scenarios for isolated spiral galaxies \citep[see a review by][]{Dobbs2014}. One of them is a classical density-wave scenario, where the spiral arms are considered to be long-lived and rigidly rotating density wave features \citep{Lin1964,Lin1966,BertinLin1996}. The other one is a transient dynamic spiral arm scenario, which is commonly seen in $N$-body simulations of disk galaxies. In this scenario, the spiral arm is short-lived, transient and recurrent \citep[][]{Sellwood1984,Fujii2011,D'Onghia2013}, and the arm is co-rotating and winding with the stars at every radius \citep[][]{Wada2011,Grand2012b,Baba2013}. This is also the case for barred spiral galaxy simulations \citep[][]{Baba2009,Grand2012a,Baba2015a}. In the density-wave scenario, the stellar arm is expected to have different degrees of offset from the gas arm at different radius \citep[e.g.][]{Fujimoto1968,Roberts1969,Gittins2004}, and hence the gas and spiral arms have different pitch angles \citep[e.g.][]{Pour-Imani2016}. On the other hand, the dynamic spiral arm scenario predicts no systematic offset of the gas and stellar arms, because they are co-rotating with each other \citep[][]{Grand2012b,Kawata2014,Baba2015}. 

This paper investigates the stellar overdensity in the Local arm and the offset of the stellar arm from the gas arm identified with the HMSFRs. Section~\ref{sec:data} describes our selection of the 1~Gyr age stellar population and discusses the distance range where the selected population shows reasonable completeness. Section~\ref{sec:res} shows our results. Summary and discussion are provided in Section~\ref{sec:sum}.


\section{Data} 
\label{sec:data}

The data used in this paper are described in this section. We first explain how we selected our sample of stars to analyze the surface stellar density map around the Local arm in Section~\ref{sec:stars}. Then, in Section~\ref{sec:VLBI} we describe the maser sources associated with HMSFRs, which are used to define the location of the Local arm for the star forming regions, i.e. gas. We assumed the solar radius of $R_0=8.34$~kpc \citep{Reid2014} in this paper. 

\subsection{Stellar Data}
\label{sec:stars}

In this paper we focus on the stellar population with the age of around 1~Gyr, and measure the surface density map to test if there is any stellar overdensity in the Local arm. Although the age of 1~Gyr is a relatively young age, they are older than the populations which are used to identify the Local arm in the past, and we consider that it is old enough to represent the stellar mass distribution of the Galactic thin disk stars. Also, the relatively young stellar population was chosen, because the color and magnitude ranges of their turn-off stars are more isolated in the HR diagram. In this paper, we focus on the region of the Galactic longitude between $l=90^{\circ}$ and $l=270^{\circ}$, because the dust extinction is less severe and a clear excess of the HMSFRs is observed and identified as the Local arm \citep{Reid2014} in this longitude range. We select the 1~Gyr stellar population from the HR diagram in near-infrared, to minimize the dust extinction. We cross-matched {\it Gaia}~DR2 with the 2MASS PSC \citep{Skrutskie2006}, using the official $Gaia$ DR2-2MASS cross-match best neighbor table \citep{Marrese2018}. For the 2MASS PSC \citep{Skrutskie2006} we select the sample whose near-infrared photometric quality flag of $K_s$ band is at least ``$A$'', which means scan signal to noise ratio greater than 10.  \citet{Bennett+Bovy2019} argues that the $Gaia$~DR2-2MASS cross-matched sample is complete within $7<G<17$~mag with conservative estimates. Following their approach, we select the {\it Gaia}~DR2 sample within $7<G<17$~mag. 
 The total number of this sample is 39,253,853 ($Gaia$~DR2-2MASS sample).
 
\begin{figure}[t]

\includegraphics[height=6.8cm,keepaspectratio]{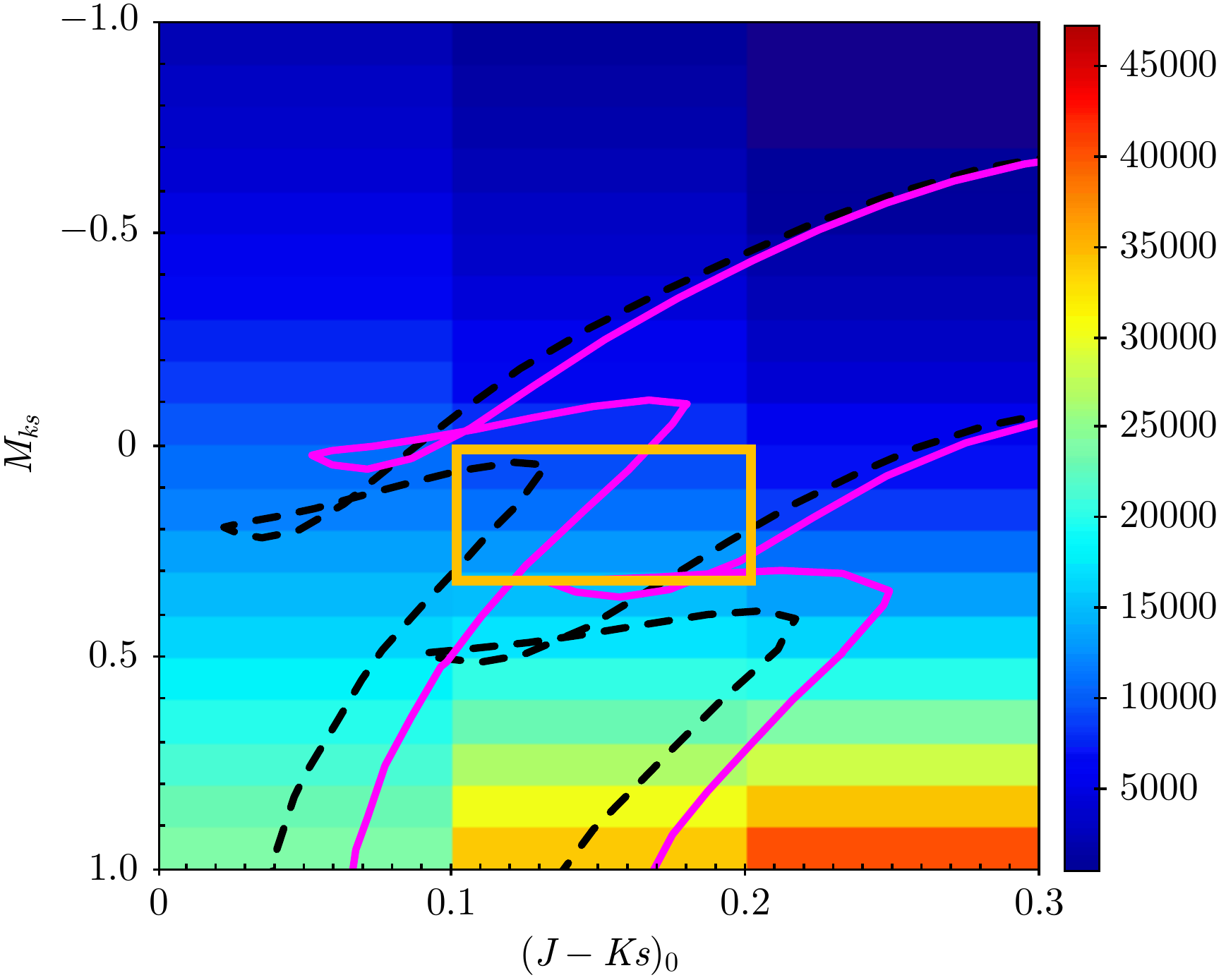}
\caption{\footnotesize
Color-magnitude distribution of $M_{K_s}$ vs. $(J-K_s)_0$ for our sample after cross-matching $Gaia$~DR2 and 2MASS data. The bin size is 0.1 $\times$ 0.1 mag and the color indicates number of stars per bin as shown with the color bar at the right. The left and right magenta (black dashed) lines indicate the PARSEC$+$COLIBRI isochrone with an age of 1 and 1.5 Gyr, respectively, with the solar metallicity (a metallicity of $Z= 0.0096$). The solid box represents the color-magnitude ranges for selecting our 1 Gyr age population. \label{fig:hr}}
\end{figure}

From the $Gaia$~DR2-2MASS sample, we further select stars whose accurate measurement of the parallax is available in {\it Gaia}~DR2 \citep{Gaia2018a} with a relative parallax uncertainty of $\pi/\sigma_{\pi}>5$ (where $\pi$ and $\sigma_\pi$ are the parallax and its uncertainty, respectively). Because we are interested in the surface density of the disk stars, we select stars within $|z|<0.3$ kpc, where $|z|$ is defined as a vertical position with respect to the Sun's vertical position, $z_{\sun}$.
When we evaluate $|z|$, we simply assume $d=1/\pi$ without taking into account the uncertainty in parallax. To estimate the three-dimensional Galactic dust extinction correction, we employed {\tt MWDUST}\footnote{\url{https://github.com/jobovy/mwdust}} \citep{Bovy2016}. This allows us to obtain extinction corrected color, $(J-K_s)_0$, and absolute magnitude, $M_{Ks}$, i.e. the HR diagram as shown in Figure~\ref{fig:hr}.
  
We then select the stars within $0.1 \leq (J-K_s)_0 \leq 0.2$ and $0.0 \leq M_{Ks} \leq 0.3$ in the HR diagram as our sample for 1~Gyr stellar population, which leaves 33,718 stars. This region in the HR diagram corresponds to the box area highlighted in Figure~\ref{fig:hr}. Figure~\ref{fig:hr} also shows the track of PARSEC$+$COLIBRI isochrones \citep{Bressan2012, Marigo2017} with ages of 1 and 1.5 Gyr with the solar metallicity ($Z_\odot= 0.0152$) and a lower metallicity of $Z=0.0096$. The figure indicates that our selected region in the HR diagram corresponds to turn-off stars whose age is between about 1 and 1.5 Gyr with the metallicity expected in the disk stars outside of the solar radius, where we focus on in this paper. There could be some contamination of blue horizontal branch stars of the Galactic halo stars in this color and magnitude range. We analyzed the vertical distribution of our selected stars, and confirmed that they are mostly confined within $|z|<0.2$~kpc. 
We also confirmed the stellar density drops rapidly with $|z|$. Hence, we consider that in the volume we study in this paper the thin disk stars are dominant, and the contamination of blue horizontal branch stars is negligible. 
 
We found that our sample is estimated to be complete in an acceptably high level within 0.2 $\leq$ $d$ $\leq$ 1.33~kpc. This corresponds to the projected distance in the disk plane, $d_{xy}$, of $0.2<d_{xy}<1.3$~kpc, when the sample is limited within $|z|<0.3$~kpc as mentioned above. We exclude the stars within $d_{xy}<0.2$~kpc, because $Gaia$~DR2 is incomplete for nearby bright stars with $G<7$~mag. We also exclude the stars at $d_{xy}>1.3$~kpc, because the completeness drops at the farther distance. Hence, our final sample used in this paper is limited within $0.2<d_{xy}<1.3$~kpc and $|z|<0.3$~kpc. This leaves our final sample of 4,654 stars.

 
\begin{figure}[htbp]
\includegraphics[width=8.6cm,keepaspectratio]{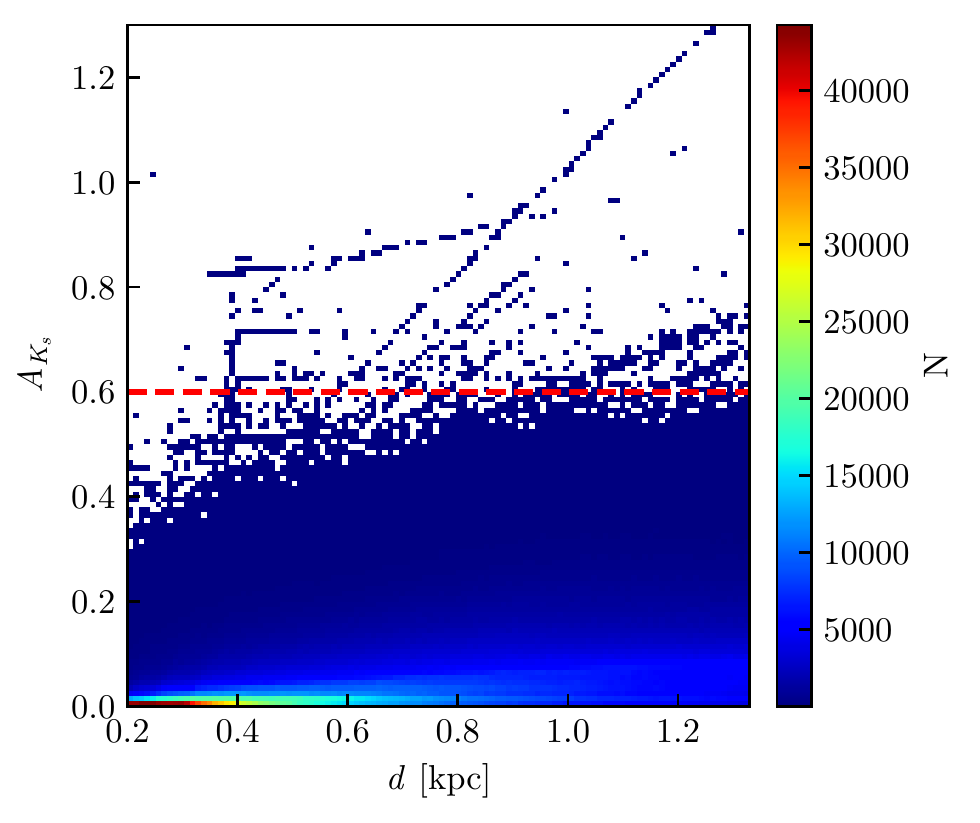}
\caption{\footnotesize
The estimated $K_s$-band extinction, $A_{K_s}$ for our selected stars as a function of the distance, $d$. 
The color indicates number of sources per bin as indicated in the color bar at right. The red horizontal dotted lines indicate $A_{K_s} = 0.6$~mag. \label{fig:reddening}}
\end{figure}

We evaluated that our sample is complete in an acceptably high level within the distance ($d$) between 0.2 and 1.33~kpc as follows.
Figure~\ref{fig:reddening} shows the extinction, $A_{K_s}$, in our selected Galactic longitude region, i.e. $90^\circ \leq l \leq 270^\circ$, within $|z|<0.3$~kpc from the Sun and the distance within $0.2 \leq d \leq 1.33$~kpc, using {\tt MWDUST}. We found that almost all the sample have $A_{K_s}<0.6$~mag. Then, our sampled absolute magnitude range of $0.0 \leq M_{Ks} \leq 0.3$~mag corresponds to $6.5 \leq K_s \leq 11.5$~mag at $0.2 \leq d \leq 1.33$~kpc with the maximum extinction of $A_{K_s}<0.6$~mag. The brightest limit of 6.5~mag corresponds to the apparent magnitude of $M_{K_s}=0$~mag at the minimum distance of $d=0.2$~kpc and the faintest limit corresponds to the apparent magnitude of $M_{Ks}=0.3$~mag at $d=1.33$~kpc plus the maximum extinction of $A_{K_s}=0.6$~mag. We found that our sample within the square region of Figure~\ref{fig:hr}, i.e. $0.1 \leq (J-K_s)_0 \leq 0.2$~mag and $0.0 \leq M_{Ks} \leq 0.3$~mag, are within $0.5<G-K_s<5.5$. This means that for our sample $6.5 \leq K_s \leq 11.5$~mag corresponds to $7<G<17$~mag. As discussed above, according to \citet{Bennett+Bovy2019} that $Gaia$~DR2$-$2MASS cross-matched sample is complete within $7<G<17$~mag. 

However, our final sample is additionally limited to the stars with $\pi/\sigma_{\pi}>5$. Hence, we compare our final sample with the $Gaia$~DR2-2MASS sample in our selected volume and color-magnitude range. To this end, we made a comparison sample, Group~C, from the $Gaia$~DR2-2MASS sample, which has $7<G<17$~mag cut, by selecting the stars within $0.1 \leq (J-K_s)_0 \leq 0.2$~mag, $0.0 \leq M_{Ks} \leq 0.3$~mag, $|z|<0.3$~kpc and $0.2<d_{xy}<1.33$~kpc, using their distance of $d=1/\pi$, regardless of their uncertainty. Parallax uncertainties of Group~C can be large. Therefore, there are contamination from the stars whose true distance, their absolute magnitude or color is not within the selected range. Also, Group~C must be missing some stars whose true distance, color and magnitude are within the selected range of our final sample, but not in Group C, because of their error in parallax. Here, we consider that these can compensate each other in some degree, and Group C is close to being a reasonable representative complete sample to be compared with our final sample for simplicity. We obtain 4,798 stars in Group~C and 4,654stars satisfy $\pi/\sigma_{\pi}>5$.
This leads to $97\%$ stars in Group~C being in our final sample, and the additional cut of $\pi/\sigma_{\pi}>5$ does not reduce the sample fraction significantly. 
Hence, our final sample is considered to be a reasonably representative sample to estimate the density distribution of our selected population of stars. 

Note that here we used the cataloged values of parallax, color and magnitude in $Gaia$~DR2 and 2MASS without taking into account the uncertainty. In addition, 3D dust extinction is still uncertain even in this area of relatively near the Sun, but we simply use {\tt mwdust} to correct the dust extinction, for simplicity. These errors in these measurements and uncertainties in the dust extinction affect which stars are included in our color-magnitude range or within the chosen spatial range of distance and the height. In this paper, we assume that these errors affect in both ways in increasing and decreasing the sample compared to the true sample. These compensate each other, and the final results are less affected by these uncertainties. Still, more statistical test is required to properly assess the effects of these errors. We postpone such statistical study to a future paper, but this paper provides more qualitative indications from the selected sample which are chosen to be a representative sample for our selected stellar population. 

\newpage
\subsection{High Mass Star Forming Regions and the Local arm} \label{sec:VLBI}

 As discussed in Section~\ref{sec:intro}, the Local arm is currently identified by the star forming regions, young OB stars and young open clusters \citep[e.g.,][]{Xu2018}. To define the location of the Local arm where the stars are forming, we use HMSFRs as shown in \citet{Xu2016}, to compare the stellar density distribution from our selected 
{\it Gaia}~DR2 stars in the previous section. 
To define the location of the Local arm from the star forming regions, using the model described in \citet{Reid2014}, we fit the distribution of HMSFRs with a following logarithmic spiral-arm model, 
\begin{equation}
\label{eq:arm}
\ln{(R/R_{\rm ref})} = -(\beta-\beta_{\rm ref})\tan \psi, 
\end{equation}
where $R_{\rm ref}$ is a reference Galactocentric radius, 
$\beta_{\rm ref}$ is a reference azimuthal angle and $\psi$ is a pitch angle. The zero point of the Galactocentric azimuthal angle, $\beta$, is defined as a line toward the Sun from the Galactic center and the angle increases toward the direction of the Galactic rotation. $\beta_{\rm ref}$ was set near the midpoint of the azimuthal angles for the Local arm HMSFR sources in \citet{Xu2016}.

Because in this paper we focus on the stellar density distribution in the second and third Galactic quadrants ($90^{\circ} \leq l \leq 270^{\circ}$), we apply the fitting with equation~(\ref{eq:arm}) to the 12 Local arm HMSFR sources within $90^{\circ} \leq l \leq 270^{\circ}$. Then, we obtained ($R_{\rm ref}$, $\beta_{\rm ref}$) = (8.87 $\pm 0.13$ kpc, 1.4$^\circ$) and $\psi$ = 13.1$^\circ \pm 7.5^\circ$. The arm's width, $a_{w}$, defined as the 1$\sigma$ scatter in the sources perpendicular to the fitted arm position, is 0.19 kpc. The location of the Local arm identified with HMSFRs (HMSFRs-defined Local arm, hereafter) is shown with the solid line in Figure~\ref{fig:density}, and the dashed lines show the width of the arm. Note that two HMSFRs are outside of the region shown in Figure~\ref{fig:density}, where only 10 HMSFRs are seen. Also note that this location of the HMSFRs-defined Local arm is different and has a significantly larger pitch angle than the Local-arm identified in \citet{Xu2016}, because \citet{Xu2016} included the Local arm HMSFRs sources in the lower Galactic longitude, $70^{\circ} \leq l \leq 270^{\circ}$. Our result provide better fit in the region where we are interested in this paper. The spiral arm may be segmented and different part of the arm may have different pitch angles \citep[e.g.][]{Honig2015}. Therefore, we use our new fit of the HMSFRs-defined Local arm in the region of our interest in this paper. 

\citet{Xu2016,Xu+Hou+Wu18,Xu2018} identified a minor segment situated between the Local and the Sagittarius arms in the first quadrant. We confirmed that the extrapolation of the minor segment does not come close to the HMSFRs-defined Local arm. Hence, we do not consider the minor segment is related to the HMSFRs-defined Local arm in the region we focus in this paper. 


\section{Results} 
\label{sec:res}

\begin{figure*}[htbp]
\centering
\includegraphics[width=18cm,keepaspectratio]{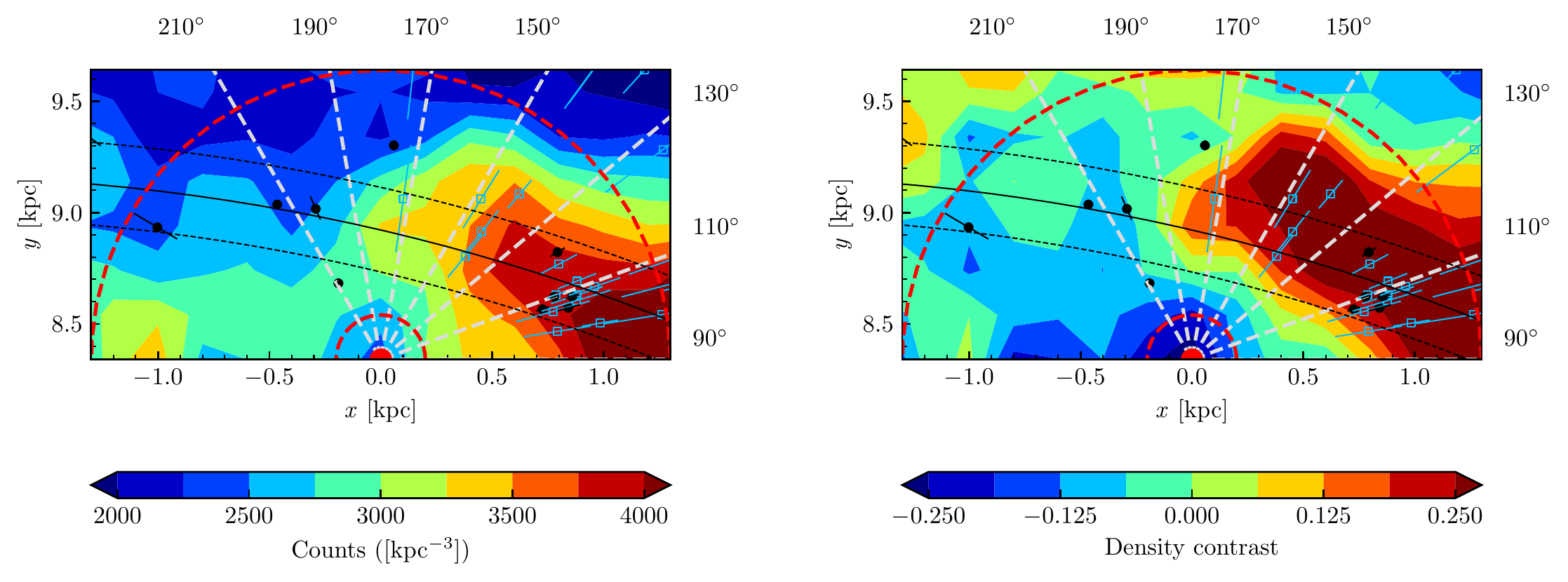}
\caption{\footnotesize
The smoothed density distribution of our selected stars as 1 Gyr stellar populations (Left) and the distribution after divided by an exponential profile with the scale length $r_d = 2.5$ kpc \citep[e.g.][]{Bland2016} (Right). The Sun is located at $(x, y) = (0, 8.34)$~kpc. $x$-axis is the direction of the Galactic rotation, and $y$-axis is the direction from the Galactic center to the Sun. The location of the Local arm defined with HMSFRs is highlighted by a solid black line and the black dashed lines indicate the width of the arm defined in Section~\ref{sec:VLBI}. The  inner and outer red dashed lines correspond to the distances from the Sun $d_{xy}=0.2$ and 1.3~kpc, and we consider that the completeness of our sample is reasonably high in the area between these lines (see Section~\ref{sec:stars}). Filled black circles with error bars show the location of HMSFRs, and the error bars correspond to the distance uncertainties. Open blue squares with error bars indicate the locations of $\rm{H_{II}}$ regions in $90^{\circ} \leq l \leq 190^{\circ}$ from \citet{Foster2015}. \vspace{1pt}}
\label{fig:density}
\end{figure*}

Using the sample of stars selected as 1~Gyr old population as described in Section~\ref{sec:stars}, we analyzed the surface density distribution of 1~Gyr old stars. The smoothed surface density distribution is shown in the left panel of Figure~\ref{fig:density}. Here, we define $x$-axis as a direction of the rotation from the Sun with the Sun's location at $x=0$~kpc, and the $y$-axis is the direction from the Galactic center to the Sun whose location is $y=r_0=8.34$~kpc. The red inner and outer dashed half-circles indicate the distance from the Sun of $d_{xy}=0.2$ and 1.3~kpc, respectively. As discussed in Section~\ref{sec:stars}, the completeness of our sample drops rapidly inside of the inner red dashed circle or outside of the outer red dashed circle. Hence, we trust the density map only between these two red dashed half-circles. 
The solid line shows the location of the Local arm defined with the HMSFRs-defined Local arm in Section~\ref{sec:VLBI}.
Interestingly, there is a clear stellar overdensity of the 1 Gyr old stars at a similar location to the HMSFRs-defined Local arm at $90^{\circ} \leq l \leq 190^{\circ}$, or slightly outside of the HMSFRs-defined Local arm at a larger Galactic longitude at $l\geq130^{\circ}$. The most significant stellar overdensity is seen between $l=90^\circ$ and $l=110^{\circ}$. 
The stellar overdensity looks extended along the HMSFRs-defined Local arm from $l=90^{\circ}$ to $l=190^{\circ}$ at least. In the region of the larger Galactic longitude than $l=190^{\circ}$, although there are 
HMSFRs, 
and the HMSFRs-defined Local arms extends continuously as shown in the black solid line and the dashed lines, the extension of the stellar overdensity is not clear within our distance limit. 


To take into account the mean stellar density decrease with the increasing Galactocentric radius, the right panel of Figure~\ref{fig:density} shows the smoothed density distribution of 1 Gyr stars after divided by an exponential density profile with the scale length of $r_d = 2.5$ kpc \citep[e.g.,][]{Bland2016}. The arm-like structure of the stellar overdensity in $90^{\circ} \leq l \leq 190^{\circ}$ as seen in the left panel of Figure~\ref{fig:density} still exists after taking into account the decrease in the stellar density at the outer radii.

\begin{figure*}[htbp]
\centering
\includegraphics[width=15cm,keepaspectratio]{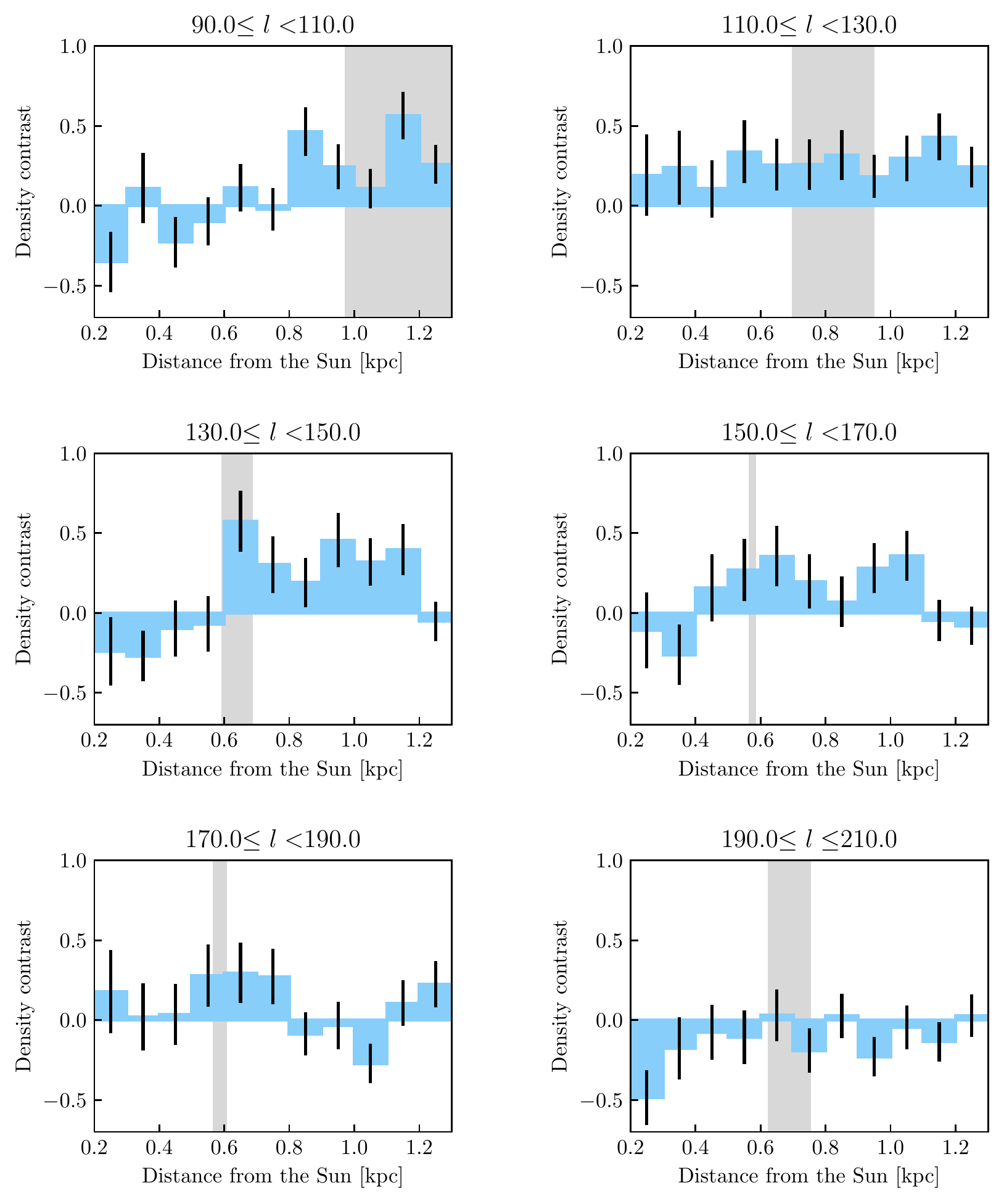}
\caption{\footnotesize 
 The stellar density contrast as a function of the distance (blue histogram) at the different Galactic longitude ranges as indicated at the top of each panel. The stellar density contrast shown in these panels are the surface density of our sample after divided by an exponential profile as the right panel of Fig.~\ref{fig:density}. The vertical error bars show the uncertainties evaluated by 500 bootstrap sampling. The vertical grey area indicates the distance range of the HMSFRs-defined Local arm as highlighted with the solid black line in Figure~\ref{fig:density} in the corresponding Galactic longitude ranges. 
 \label{fig:density_boot}}
\end{figure*}

To quantify the significance of this overdensity, we compute the stellar density after divided by the exponential law as a function of the distance from the Sun in the different longitudinal regions between $l=90^{\circ}$ and $l=210^{\circ}$. The results are shown in Figure~\ref{fig:density_boot}.  We evaluated the uncertainty of the density in each distance bin by taking the dispersion of the density in each bin measured from 500 bootstrap realization of the sample. Within $90^{\circ} \leq l \leq 110^{\circ}$ the stellar density increases with the distance from the Sun, and the density peak is seen at close to our distance limit of $d_{xy}=1.3$~kpc, as seen in Figure~\ref{fig:density}.  Although it is not very clear at $110^{\circ} \leq l \leq 130^{\circ}$, at $130^{\circ}\leq l \leq 190^{\circ}$, we can see a more clear peak of density within our distance limit, and the density decreases with the increasing distance after passing the density peak. This peak corresponds to the arm-like overdensity seen in Figure~\ref{fig:density}, and the significance of the overdensity is about $2\sigma$, when comparing the highest density peak and the uncertainty especially at $130^{\circ}\leq l \leq 170^{\circ}$. Hence, we think that the stellar overdensity seen in Figure~\ref{fig:density} is very likely a real structure. 

 Figure~\ref{fig:density_boot} also confirms that at $l \geq 190^{\circ}$ the overdensity is not very clear. The histograms of the density distribution as a function of distance in $190^{\circ} \leq l \leq 210^{\circ}$ show a hint of the overdensity, but it is not statistically significant. Hence, as seen in Figure~\ref{fig:density}, at $l \geq 190^{\circ}$ there is no significant stellar component corresponds to the HMSFRs-defined Local arm or they are farther than the distance we can confidently analyze the stellar density. 
 
The vertical grey area in Fig.~\ref{fig:density_boot} shows the distance range of the HMSFRs-defined Local arm in the corresponding Galactic longitude range in each panel. Interestingly, at $130^{\circ} \leq l \leq 190^{\circ}$ the location of the stellar Local arm overdensity is slightly outside of the HMSFRs-defined Local arm at a larger Galactic longitude. 
On the other hand, at lower Galactic longitude, $110^{\circ} \leq l \leq 130^{\circ}$ the HMSFRs-defined local arm is located at the center of the stellar overdensity, although the width of the arm overdensity is not very clear. This trend is also seen in the right panel of Figure~\ref{fig:density}. If this is true, the stellar Local arm we identified has a slightly larger pitch angle than the HMSFRs-defined Local arm. We will discuss the implication of this result in Section~\ref{sec:sum}.

\section{Summary and Discussion} \label{sec:sum}

Taking advantage of the precise measurements of parallax for a large number of stars recently provided by {\it Gaia} DR2, we analyzed surface stellar density map for a relatively old ($\sim1$~Gyr) stellar populations of the thin disk stars between $90^{\circ} \leq l \leq 270^{\circ}$. We identified the 1~Gyr population from a carefully chosen range of the color and magnitude in the near-infrared bands, after cross-matching {\it Gaia}~DR2 and 2MASS. We evaluated that our sample is reasonably complete within the distance between 0.2 and 1.3~kpc. We found a marginally significant arm-like stellar overdensity close to the Local arm identified with the HMSFRs especially in the region of $90^{\circ} \leq l \leq 190^{\circ}$. At $l\geq 190^{\circ}$ we could not find a significant stellar overdensity. At $90^{\circ} \leq l \leq 190^{\circ}$ the identified stellar Local arm is located similar region to the HMSFRs-defined Local arm. Our finding indicates that the Local arm is not a minor arm with only the gas and star forming clouds, but a significant stellar overdensity is associated, too. 
 
Interestingly, at $130^{\circ} \leq l \leq 190^{\circ}$ the identified stellar Local arm is located slightly outside of the HMSFRs-defined Local arm, while at lower Galactic longitude of $90^{\circ} \leq l \leq 130^{\circ}$ the stellar Local arm is co-located with the HMSFRs-defined Local arm. This indicates that the pitch angle of the stellar arm is slightly larger than the HMSFRs-defined arm, and there is an offset between HMSFRs-defined (i.e. gas) and stellar arms especially at the larger Galactic longitude, $130^{\circ} \leq l \leq 190^{\circ}$.
The offset and different pitch angles between the stellar and gas spiral arms are consistent with what is expected from a classical density-wave and its galactic shock theory \citep[e.g.,][]{Roberts1969}. Hydrodynamic simulations with the rigidly rotating spiral arm potentials also consistently show that the pitch angle of the gas arm is smaller than that of the (stellar) spiral arm \citep{Gittins2004,Baba2015}. However, we note that the pitch angle of the stellar Local arm is larger compared to the other major spiral arms like the Perseus arm and the Scutum-Centaurus arm \citep[e.g.][]{Reid2014}. This could be an issue for a classical density-wave theory where a constant pitch angle is expected in the different spiral arms at least at the same radius. More complicatedly, \citet{Vallee2018ApJ} found an offset between the CO arm and the HMSFRs-defined arm in the Perseus arm, where the CO spiral arm (earlier phase of gas spiral arm) shows a larger pitch angle than the HMSFRs-defined spiral arm. How this can be compared with the offset between the stellar arm and the HMSFRs-defined arm is not a trivial question and \citet{Pour-Imani2016} argues that there are two scenarios of the offset of stellar and gas spiral arms in the density-wave scenario. It is required to further study the offset between different arm tracers in the different spiral arms in the Milky Way. More clear theoretical predictions in the density-wave scenario are also necessary. 

In Figure~\ref{fig:density}, we also show the locations of the $\rm{H_{II}}$ regions from \citet{Foster2015} who measured the distance to the $\rm{H_{II}}$ regions within $90^{\circ} \leq l \leq 190^{\circ}$. The $\rm{H_{II}}$ regions are also located in the similar region to the stellar overdensity we identified. Interestingly, the right panel of Figure~\ref{fig:density}, which shows the stellar overdensity after taking into account the mean stellar density decrease with the increasing Galactocentric radius, tentatively shows that the $\rm{H_{II}}$ regions seem to be located between the identified stellar arm and the HMSFRs-defined Local arm. Admittedly, this is a quite tentative trend with very low number statistics. However, if this is confirmed, because the $\rm{H_{II}}$ regions are considered to be the star formation tracer phase later than HMSFRs, the offset between HMSFRs, the $\rm{H_{II}}$ regions and the stellar arm would provide the strong support for the density-wave scenario. 
  
\citet{Lepine2017} suggested that the Local arm is caused by the trapped orbits at the co-rotation resonance of the major spiral arms of the Perseus and the Sagittarius-Carina arms. Our stellar overdensity in the similar location to the Local arm traced by HMSFRs and $\rm{H_{II}}$ regions does not contradict with this scenario. However, if the offset between these different populations is confirmed to be true, this scenario may be difficult to explain such offset and different pitch angles for different populations. The stellar Local arm identified in this paper encourages further quantitative comparison between the model and the observed features.

  
The offset between the gas and stellar arms is also difficult to be explained with the dynamic spiral arm scenario, where no ``systematic'' offset between the stellar and star forming arms are expected \citep{Grand2012b,Baba2015}. On the other hand, the large pitch angle of the Local arm is consistent with the currently forming spiral arm for the dynamic spiral arm scenario \citep{Baba2013,Grand2013}. Also, the recent observations of the converging velocity field around the Local arm \citep{Liu2017} is consistent with the ongoing formation of the Local arm. This converging velocity field is different from the diverging velocity field observed around the Perseus arm \citep{Baba2018,Tchernyshyov+2018}. The dynamic spiral arm scenario can explain these differences in the velocity field and pitch angles between the Perseus arm and the Local arm, if the Perseus arm is in a disrupting phase, while the Local arm is in a building up phase \citep{Baba2013,Grand2014,Baba2018}. However, the dynamic spiral arm scenario has to be able to explain the significant offset found in this paper. The significant external perturbation \citep[but see also][for an alternative scenario]{Micthchenko2019}, which is suggested to explain the Galactic disk in-plane and vertical motions found in {\it Gaia}~DR2 \citep{Kawata2018,Antoja2018,Bland-Hawthorn2019}, may affect the origin of the spiral arm in the Milky Way \citep[][]{Laporte2019}, and may be able to explain this offset. Further modelling of the spiral arms including the external perturbations of the Galactic disk is necessary to further understand the formation mechanism of the spiral arm. 
  
Unfortunately, the edge of the stellar Local spiral arm is close to our distance limit. Also, although we carefully take into account the completeness of the {\it Gaia}~DR2 data, further studies with better data and also taking into account the selection function \citep[e.g.][]{Bovy2017} are required to accurately map the Local arm and the other spiral arm at the farther distances, and answer this long-standing challenge of understanding the origin of the spiral arms in the Milky Way. The result of this paper provides a new and complex picture of the Local arm,
and encourages such further work. The expected parallax accuracy for the fainter stars in the next {\it Gaia} data releases will certainly help, to map the stellar density structures for the different age populations. 
Ultimately the near-infrared astrometry mission like the {\it small-JASMINE} \citep{Gouda2012} and ultimately the mission like {\it Gaia NIR} concept \citep{Hobbs2016} would be required to answer the structure and origin of the spiral arms. \newline \\

\acknowledgments
We are grateful to the anonymous referee for constructive comments that have improved the paper. We acknowledge Dr. Mark J. Reid for allowing us to use his MCMC program, utilized to define a locus of the Local spiral-arm, traced by HMSFRs. DK thanks the generous support and hospitality of the National Astronomical Observatory of Japan and the Kavli Institute for Theoretical Physics (KITP) in Santa Barbara during the 'Dynamical Models For Stars and Gas in Galaxies in the Gaia Era' program. KITP is supported in part by the National Science Foundation under Grant No. NSF PHY-1748958. DK also acknowledges the support of the UK's Science \& Technology Facilities Council (STFC Grant ST/N000811/1).
JB is supported by the Japan Society for the Promotion of Science (JSPS) Grant-in-Aid for Scientific Research (C) Grant Number 18K03711.
NM and JB are supported by the JSPS Grant-in-Aid for Scientific Research (B) Grant Number 18H01248.
Data analysis was carried out on the Multi-wavelength Data Analysis System operated by the Astronomy Data Center (ADC), National Astronomical Observatory of Japan. This work has made use of data from the European Space Agency (ESA) mission {\it Gaia} (\url{https://www.cosmos.esa.int/gaia}), processed by the {\it Gaia} Data Processing and Analysis Consortium (DPAC; \url{https://www.cosmos.esa.int/web/gaia/dpac/consortium}). Funding for the DPAC has been provided by national institutions, in particular the institutions participating in the {\it Gaia} Multilateral Agreement. This publication makes use of data products from the Two Micron All Sky Survey, which is a joint project of the University of Massachusetts and the Infrared Processing and Analysis Center/California Institute of Technology, funded by the National Aeronautics and Space Administration and the National Science Foundation. We are indebted to NASA's \href{http://adsabs.harvard.edu}{ADS} for its magnificent literature and bibliography serving.


\end{document}